# Self-assembly of colloidal particles from evaporating droplets: role of DLVO interactions and proposition of a phase diagram

Rajneesh Bhardwaj[a], Xiaohua Fang[b], Ponisseril Somasundaran[b] and Daniel Attinger[a*]

[a]Laboratory for Microscale Transport Phenomena,

Department of Mechanical Engineering,

[b]Langmuir Center of Colloids and Interfaces,

Department of Earth and Environmental Engineering,

Columbia University, New York, NY 10027

*Corresponding author. Tel: +1-212-854-2841

E-mail address: da2203@columbia.edu (D. Attinger)

**Abstract**

The shape of deposits obtained from drying drops containing colloidal particles matters for technologies such as inkjet printing, microelectronics and bioassay manufacturing. In this work, the formation of deposits during the drying of nanoliter drops containing colloidal particles is investigated experimentally with microscopy and profilometry, and theoretically with an in-house finite-element code. The system studied involves aqueous drops containing titania nanoparticles evaporating on a glass substrate. Deposit shapes from spotted drops at different pH values are measured using a laser profilometer. Our results show that the pH of the solution influences the dried deposit pattern, which can be ring-like or more uniform. The transition between these patterns is explained by considering how DLVO interactions such as the electrostatic and van der Waals forces modify the particle deposition process. Also a phase diagram is proposed to describe how the shape of a colloidal deposit results from the competition between three flow patterns: a radial flow driven by evaporation at the wetting line, a Marangoni recirculating flow driven by surface tension gradients, and the transport of particles towards the substrate driven by DLVO interactions. This phase diagram explains three types of deposits commonly observed experimentally, such as a peripheral ring, a small central bump, or a uniform layer. Simulations and experiments are found in very good agreement.

# Nomenclature

| | |
|---|---|
| $a$ | arbitrary constant |
| $A$ | Hamaker constant (= $2.43 \times 10^{-20}$ J for water [1]) |
| $C$ | concentration [mol/L] |
| $d$ | diameter [m] |
| $D_{pl}$ | diffusion coefficient of particles in liquid [m²/s] |
| $e$ | electronic unit charge [$1.6 \times 10^{-19}$ C] |
| $F$ | force between a particle and substrate [N], Faraday constant (96500 C/mol) |
| $g$ | gravitational acceleration [9.81 m/s²] |
| $H$ | Relative humidity [-] |
| $I$ | ionic strength of solution [mol/L] |
| $j$ | evaporative mass flux [kg m$^{-2}$s$^{-1}$] |
| $k_B$ | Boltzmann constant [$1.38 \times 10^{-23}$ J K$^{-1}$] |
| $m$ | mass of particles [kg] |
| $n$ | number concentration [molecules/m³] |
| **n** | unit normal vector, **n** = (**n**$_r$ · **n**$_z$) |
| $N_A$ | Avogadro number [$6.023 \times 10^{23}$ molecules/mol] |
| $r$ | radial coordinate [m] |
| $S$ | specific surface area [m²/kg] |
| $t$ | time [s] |
| $T$ | absolute temperature [K] |
| **v** | velocity vector, **v** = ($u$, $v$) |
| $v$ | axial velocity [m s$^{-1}$] |
| $V$ | Volume of drop [nL], magnitude of the velocity vector **v** [m s$^{-1}$] |
| $X$ | concentration of particles [kg of particles/kg of solution] |
| $z$ | axial coordinate [m] |

**Greek letters**

| | |
|---|---|
| $\alpha_{rtd}$ | retardation factor for van der Waals force |
| $\beta$ | gradient of surface tension with temperature [N m$^{-1}$ K$^{-1}$] |
| $\gamma$ | function of surface potential [-], surface tension [N m$^{-1}$] |
| $\phi$ | wetting angle of the drop [-] |
| $\varepsilon$ | dielectric constant of the water [78.54] |
| $\varepsilon_0$ | vacuum permittivity [8.85×10$^{-12}$ C$^2$ J$^{-1}$ m$^{-1}$] |
| $\kappa^{-1}$ | Debye screening length [m] |
| $\psi$ | surface potential [V] |
| $\mu$ | dynamic viscosity [Pa s] |
| $\rho$ | density of drop liquid [kg/m$^3$] |
| $\sigma$ | surface charge density [Cm$^{-2}$] |

**Subscripts**

| | |
|---|---|
| amb | ambient |
| drag | hydrodynamic drag |
| DLVO | Derjaguin-Landau-Verwey-Overbeek theory |
| el | electrostatic |
| + | attraction |
| - | repulsion |
| $i$ | initial |
| max | maximum value |
| Ma | Marangoni |
| *l* | liquid |
| *p* | particle |
| rad | radial |
| rec | receding, depinning |
| s | substrate |
| vdW | van der Waals force between a particle and substrate |
| vdWp | van der Waals force between two particles |
| *z* | axial |

# 1   Introduction

In biology, spotting and evaporation of drops containing colloidal particles is used for depositing and organizing biological materials such as proteins and DNA [2-8]. Colloidal deposition and crystallization [9-15] can also be used to manufacture micro- and nanowires [16-17], nanocrystals [18], cosmetics and explosive crystalline layers [19]. Figure 1 shows that the patterns left by an evaporating drop containing colloidal particles can exhibit a ring-like structure [20], a central bump [21], a uniform deposit [22-23], or complex patterns involving multiple rings, a network of polygons [24], hexagonal arrays [18] or Marangoni tongues.  This variety of patterns reflects the complex, coupled and multi-scale nature of the transport phenomena occurring during the droplet evaporation. The fluid dynamics involved in droplet evaporation is transient. It depends on the Reynolds and Weber number of the droplet impact, on the impact angle and associated interfacial deformation or break-up, on the Marangoni and wetting stresses, and on the evaporation at the free surface. Heat transfer occurs by convection inside the drop and conduction in the substrate, driven by a latent heat contribution at the evaporating free surface. Mass transfer occurs through diffusion of liquid vapor in the atmosphere, advection-diffusion of particles in the drop and long range interactions between the charged particles and substrate surfaces.

Over the last decade, theoretical and experimental efforts have been made to explain the mechanisms responsible for two of the most common deposits, the peripheral ring and the central bump. In 1997, Deegan *et al.* found that a peripheral ring deposit forms [20, 25-26] because of a strong radial flow carrying particles towards the pinned wetting line, where evaporative flux is the highest due to the wedge geometry. The formation of a central bump, i.e. a hilly accumulation with a diameter much smaller than the initial wetted diameter of the drop, was explained by Hu and Larson [23] as due to recirculation loops due to thermal Marangoni stresses along the free surface. Very recently, Ristenpart et al. [27] showed analytically that the ratio between the substrate and droplet thermal conductivities controls the direction of Marangoni convection inside an evaporating drop, with a direct effect on the deposit pattern.

For applications such as inkjet printing [28-29] and bio-assays [2-8, 30], a uniform deposit with diameter equivalent to the initial wetted diameter might be desired in place of a ring or a central bump. While the mechanisms responsible for ring and central bumps are well understood, several mechanisms and explanations have however been put forth to explain the formation of uniform deposit. For instance, Park and Moon [31] investigated the particle deposition morphologies that resulted from evaporating jetted microdroplets. By varying the chemical composition of the ink, they showed that the deposit structure can change from a ring to a uniform two-dimensional monolayer with a well-ordered hexagonal structure [31]. Sommer et al. [32-33] studied the patterns formed by hydroxyapatite particles (60 nm diameter) suspended in 10 microliter aqueous drops on a titanium disk. They explained the transition from a ring to a uniform deposit patterns as the result of competition between hydrodynamic and van der Waals forces. Yan et al. [34] observed both ordered and disordered particle assembly during ring formation depending on different surface charges and the kind of surfactants. Capillary forces have also been mentioned as responsible for controlling the deposit shape: Andreeva et al. [35] reported ring and uniform patterns formation for aqueous drops on a respectively hydrophilic and hydrophobic surface, and Onoda and Somasundaran [36] showed that deposit patterns on scratched surfaces could be explained by larger capillary forces on a hydrophobic surfaces than on a hydrophilic surfaces. In the 1990s, the role of capillary forces during the evaporation of thin aqueous films containing micrometer-size polystyrene particles was investigated by the Ivanov group. They found that during the last stages of the evaporation, attractive capillary forces were responsible for the ordering of the particles [12-13, 37].

Besides experiments showing deposit patterns and their qualitative interpretations, few numerical and analytical studies have been reported to quantitatively analyze the formation of colloidal deposits. One reason is that modeling drop evaporation is challenging, given the wide range of transport phenomena, time scales and length scales. The explanation of the *coffee-ring pattern*, or ring deposition, in Deegan et al. [20, 25-26] was based on the lubrication approximation, with an analytical expression for the local evaporative flux. Hu and Larson [23] computed the deposition of PMMA particles by combining an analytical flow field with Brownian dynamics simulations. Widjaja and Harris [38] predicted particle concentrations using a finite element numerical model solving a continuum advection-diffusion equation. Using potential flow, Petsi and Burganos [39] analytically showed that the radially outward flow during the evaporation of a drop was slower on a hydrophobic surface than on a hydrophilic surface. They used this finding to explain the formation of multiple rings seen in experiments on hydrophobic surfaces [40].

This paper aims at investigating the role of Derjaguin-Landau-Verwey-Overbeek (DLVO) interactions on the shape of the patterns left by evaporated drops. To do so, we perform experiments where the pH of the drop containing colloidal particles is varied. The influence of the pH variation on the DLVO forces is quantified and shown to control the deposit from a ring like pattern to a more uniform layer (section 2 and 4.1). We also describe a numerical modeling for droplets containing colloidal particles evaporation that considers van der Waals and electrostatic forces between the particles and the substrate (section 3). Comparison between experiments and simulations in section 4.2 shows good agreement and explain qualitatively and quantitatively the observed deposit shapes. In the next section (4.3) we put forth a phase diagram describing how deposits patterns such as a uniform layer, a central bump or a peripheral ring result from the competition between three flow patterns: the radial flow driven by evaporation at the wetting line, the Marangoni recirculating flow, and the transport of particles towards the substrate driven by Derjaguin-Landau-Verwey-Overbeek (DLVO) interactions.

## 2  Experimental details

Titania particles (Sigma-Aldrich Inc, anatase nanopowder, 637254-500G) with an average diameter of 25±2 nm were dispersed in water using stirring. Solutions at five pH values from 1.4 to 11.7 were obtained by adding hydrochloric acid (0.1M) or sodium hydroxide (0.01M). The pH was measured using a digital pH meter (Accumet basic, AB15, Fisher scientific Inc). The surface area of the particles was measured as 117.7+/-1 $m^2$/g with a Quantasorb Surface Area Analyzer (Quantachrome Instruments Inc, Monosorb). To do so, particles were initially loaded in a glass sample cell, preheated at 70ºC for half an hour to remove adsorbed humidity. The surface charge density of titania particles was determined as a function of the pH by measuring the adsorbed potential-determining ions (here $H^+$) with a back titration method [41]. In this method, Titania particles are diluted into aqueous solutions at various initial values of pH at a concentration of 0.2 mg/mL. The amount $\Delta V$ of hydrochloric acid or sodium hydroxide that is equivalent for bringing the particle suspension back to the pH of the initial aqueous solution is calculated using titration curves. The number of potential determining ions $\Delta n$ being adsorbed on the particle are $\Delta n = C\Delta V$ where $C$ is the concentration of hydrochloric acid or sodium hydroxide. The surface charge density $\sigma$ can then be calculated as [42]:

$$\sigma_p = -\frac{F\Delta n}{m_p S_p} \quad \quad 1$$

where $F$ represents the Faraday Constant (96500 C/mol), $m_p$ is mass of the particles [kg], $S_p$ is the specific surface area of the particles [m$^2$/kg]. The number concentration of the counterions at infinite distance $n$ is related to the molar concentration $M_i$ of counterions by

$$n = 1000 M_i N_A \quad \quad 2$$

where $N_A$ is the Avagadro number.

Immediately after sonicating the solutions for five minutes at a power of 200W of the solutions, nanoliter drops containing colloidal particles were spotted on soda lime glass slides (Fisher Scientific Inc) using a 375 μm diameter stainless steel pin (Telechem International Inc, CA), as in [43]. All experiments were performed without controlling of the ionic strength of the solution. The glass slides were previously immersed in freshly prepared Piranha solution for 1 hour at 70°C, thoroughly washed with distilled water and then blown dry with nitrogen to remove residual water.

The evaporation of water droplets was visualized from the side using a digital camera (Pixelink, PLA 741, 1.3 Megapixel) and an Optem long-distance zoom objective. Typical time and spatial resolution were respectively 20 frames per second and 1.5 μm per pixel. The initial observed droplet volumes were typically between 4 and 6 nL. We also used an Olympus IX-71 inverted microscope to qualitatively assess the structure of the flow inside the drop: the motion of 25 nm Titania particles was recorded at frame rates of 25 frames per second [44]. After evaporation, profiles of the deposits were measured using a laser profilometer based on confocal microscopy (Keyence corporation, LT-9010-M, resolution ~ 10 nm). Note that this type of measurement method would not be appropriate for dried profiles that exhibit azimuthal instabilities, such as fingerings or the dendrimer structures in [45-46].

## 3 Theory and numerical modeling

Electrostatic and van der Waals forces between the particles and the solid substrate were estimated according to the DLVO theory. When the particles are in a solution, dissolution of ionic groups or preferential adsorption of ions both induce surface charges on their solid surfaces. For Titania nanoparticles in an acid/base solution, protonation and deprotonation events control the sign and value of surface charge density (defined as the charge per unit area on the solid). The surface potential of the particles ($\psi_p$) can be related to the surface charge density ($\sigma_p$) by the Gouy-Chapman relation [47-48]:

$$\sigma_p = \frac{2\varepsilon\varepsilon_0 k_B T}{e\kappa^{-1}} \sinh\left(\frac{e\psi_p}{2k_B T}\right) \qquad 3$$

where $\varepsilon\varepsilon_0$ is the total permittivity of the water, $k_B$ the Boltzmann constant, $T$ the absolute temperature (here 298 K), $e$ the electronic unit charge and $\kappa^{-1}$ the Debye length. We measured the surface potential ($\psi_p$) for the Titania particles in water and our results in Figure 2 show a point of zero charge at pH 5.4. For the glass substrate surface, the surface potential ($\psi_s$) is assumed to be equal to the zeta potential ($\psi_s \approx \zeta_s$), with values taken from the streaming potential measurements by Somasundaran et al. [49]. The interpolation of these values for low pH values shows a point of zero charge at 2.44. Other measurements in [50] show a similar behavior with a point of zero charge at pH 2.0. The electrostatic force between a particle and the substrate is calculated using the expression in [48]:

$$\mathbf{F}_{el} = -\frac{128\pi d_p \gamma_s \gamma_p n k_B T \kappa^{-1}}{2} \exp(-z/\kappa^{-1})\mathbf{n}_z = a\exp(-z/\kappa^{-1})\mathbf{n}_z \qquad 4$$

In the above equation, $d_p$ is the diameter of the particle [m], $n$ the number concentration of the counterions far away [molecules/m$^3$], $k_B$ the Boltzmann constant, $T$ the ambient temperature (298 K), $\kappa^{-1}$ the Debye length, $z$ is the distance between the particle and the substrate [m] and $\mathbf{n}_z$ is the unit vector normal to the substrate. The symbols $\gamma_p$ and $\gamma_s$ are functions of the surface potential of the respective particle ($\psi_p$) and substrate ($\psi_s$):

$$\gamma_p = \tanh\left(\frac{e\psi_p}{4k_B T}\right), \quad \gamma_s = \tanh\left(\frac{e\psi_s}{4k_B T}\right) \qquad 5$$

In eqs. 3 and 4, the Debye screening length $\kappa^{-1}$, thickness of diffuse electric double layer is defined as [48]:

$$\kappa^{-1} = \sqrt{\frac{\varepsilon\varepsilon_0 k_B T}{2N_A e^2 I}} \qquad 6$$

with $N_A$ the Avagadro number and $I$ the ionic strength of dispersant defined as

$$I = \frac{1}{2}\sum_{i=1}^{n} z_i^2 C_i \qquad 7$$

with $z_i$ the valence of the ions, and $C_i$ is the concentration of the dissolved ions.

The van der Waals attraction force between the substrate and a particle is given by [48, 51]:

$$\mathbf{F}_{vdW} = \frac{1}{12} A d_p^3 \frac{\alpha_{rtd}}{z^2(z+d_p)^2} \mathbf{n}_z \qquad 8$$

where $A$ is the Hamaker constant, $A = 2.43 \times 10^{-20}$ J for water [1] and $\alpha_{rtd}$ is the retardation factor for the van der Waals force, which depends on the distance between particle and substrate as described in [51]. The total DLVO force between a particle and the substrate is the algebraic sum of the electrostatic (eq. 4) and van der Waals forces (eq. 8):

$$\mathbf{F}_{DLVO} = \mathbf{F}_{el} + \mathbf{F}_{vdW} \qquad 9$$

The numerical modeling used in this paper is described in details in our earlier work [21], and is briefly described here. This model is based on a finite-element code for droplet impact and heat transfer developed by Poulikakos and co-workers in Refs. [52-57], and Attinger and co-workers in Refs. [21, 55, 58]. This model has been validated for studies involving impact and heat transfer of molten metal [52, 56] and water drops [58], and for the evaporation of drops containing colloidal particles [21]. The flow inside the droplet is assumed to be laminar and axisymmetric. All equations are expressed in a Lagrangian framework, which provides accurate modeling of free surface deformations and the associated Laplace stresses [59]. The use of a Lagrangian scheme where the nodes move with the fluid allows precise handling of free surface stresses and the full treatment of both convection and conduction heat transfer by solving the heat diffusion equation [54]. This numerical code also models the evaporative flux along the drop-air interface, thermocapillary stresses and Marangoni flow, and wetting line motion [21]. In addition, the code has a dual time-step scheme to handle multiple time scales, which range from nanoseconds for capillary waves at the liquid-air interface to several seconds for the whole evaporation [21]. The motion of particles is tracked by solving an advection-diffusion equation for the particle concentration, neglecting buoyancy. The interaction of the free surface of the drop with the growing deposit is modeled [21] using wetting angles criteria to predict the detachment of the drop liquid from the ring, or depinning. According to published results showing unexpectedly small Marangoni convection in aqueous drops [60-61], Marangoni convection is not considered in the simulations presented here. In this paper our numerical modeling is extended to consider the attractive DLVO force between the colloidal particles and the substrate, as follows.

The governing equation for the particles transport is given by [7, 62]:

$$\frac{\partial X}{\partial t} + \nabla \cdot (X\mathbf{v}) = D_{pl} \nabla^2 X \qquad \mathbf{10}$$

where $X$ is the concentration of the particles [kg of particles/kg of solution] and $D_{pl}$ is the diffusion coefficient of the particles in the drop liquid (1.7e-11 m$^2$/s for 25 nm Titania particles in water, calculated using the Stokes-Einstein equation [63]). In the above equation, the advection velocity $\mathbf{v} = \mathbf{v_{fluid}} + \mathbf{v_{DLVO}}$. The symbol $\mathbf{v}_{fluid}$ stands for the hydrodynamic velocity of the fluid. The velocity $\mathbf{v}_{DLVO}$ is estimated by balancing the DLVO force ($\mathbf{F}_{DLVO}$, eq. 9) and the hydrodynamic drag force ($\mathbf{F}_{drag}$) for a particle. In the case of an attractive DLVO force, we have

$$\mathbf{v}_{DLVO+} = -(2\mathbf{F}_{DLVO+}/6\pi\mu d_p)\mathbf{n}_z, \qquad\qquad 11$$

where the unit vector $\mathbf{n}_z$ is normal to the substrate. For the sake of numerical tractability, the following assumptions are made. The numerical code is based on the continuum assumption and does not account for collisions of individual particles.. The pH of the droplet gradually changes during the drying process since the volatilities of water and hydrochloric acid/sodium hydroxide are different. This results in a varying DLVO force with time. However, in our simulations the pH values are assumed constant with time and equal to their initial values, for the sake of computational tractability and as a first order approximation. Buoyancy, interparticle forces as well as the effects of surface heterogeneities on the solid substrate are not considered in the modeling.

## 4  Results and discussions

In section 4.1, we describe experimentally the formation of deposits during the evaporation of 4-6 nL water droplets at various pH values. The initial volume fraction of 25 nm titania nanoparticles is 2%. In the same section 4.1, the pH is shown to control the deposit shape. In section 4.2, numerical simulations are presented at two specific values of pH, resulting respectively in a ring and a uniform deposit, both in very good agreement with the experiments. Finally, in section 4.3 a phase diagram is put forth that describes how the shape of a colloidal deposit results from the competition between three characteristic flow patterns.

### 4.1  Experimental results

Figure 3 shows micrographs of deposit patterns obtained at six pH values. The deposit structure at pH 1.4 and 2.8 is a thin uniform layer with a thicker ring at the periphery. Deposit structures at pH 5.8 and 6.7 show the initial wetted area covered randomly by particle aggregates. The deposit at the highest pH value, pH 11.7, is a ring with almost no particles at the center of the deposit. Measured deposit profiles corresponding to four different pH values are shown in Figure 4. These different deposit patterns can be explained by considering the electrostatic and van der Waals forces between the particles and the substrate, the sum of which is called here DLVO force. In Table 1, calculations show an attractive DLVO force for pH $\leq$ 5.8 and a repulsive force for pH > 5.8. In the latter case, the particles are prevented from contacting the substrate and follow the general flow pattern, which is radially towards the wetting line. The associated movie [44] shows the motion of 25 nm titania particles inside the evaporating drop, which accumulate as a ring at the periphery. A radially outward flow is observed to start at 2 s and the developing ring is visible from $t$ = 2 to 6 s. This explains the ring observed at pH 11.7, with profile as in Figure 4d. When the DLVO force is attractive the particles close to the substrate are attracted to and form the layer measured at the center of the deposits in Figure 4a-b for the pH 1.4 and 2.8 cases. Since the range of the DLVO forces is of the order of the Debye length, much less than the droplet height, a significant amount of particles are not attracted to the substrate and accumulates as a ring. The associated movie [64] shows the motion of 25 nm titania particles inside the evaporating drop. A radially outward flow is observed to start at 2 s and the deposition of particles at the drop-substrate interface with the developing ring is visible from $t$ = 2 to 8 s. The main difference between the deposits at low vs. high pH is that at low pH a relatively thick uniform deposit forms, exhibiting a peripheral ring, while at high pH almost all particles are deposited in a peripheral ring (see Figure 4a, b). These deposits can therefore be explained by the competition between the two flow regimes shown at the top and at the middle of Figure 7.

The deposit structure in Figure 3 for intermediate pH values is more complex, with the agglomerated particles sparsely spread over the entire initial wetted area. This can be explained by taking into account the fact that near the point of zero charge of the particles, the DLVO force between the particles and the substrate becomes smaller than the van der Waals forces between the particles, as shown in Table 1: this causes flocculation. Another explanation would be that capillary instabilities occur in the later stages of the drying, when the drop has a film-shape, and induce these islands of particles. The associated movie [65] invalidates the second hypothesis, showing that the flocculation occurs from the very early stages of the deposition.

### 4.2 Comparison between numerical and experimental results

In this section, we describe numerical simulations performed for two specific cases at pH of 11.7 and 1.1. In first case, a 5 nL water droplet containing 2% volume fraction titania particles of 25 nm diameter evaporates at ambient temperature at pH = 11.7 ($T_{amb}$ = 25.5°C, Figure 3). Second case corresponds to the evaporation of a 4 nL water droplet at pH = 1.1 in a refrigerator with ambient temperature of 10.9°C. The particles size and concentration are same as in the first case. The values of the electrostatic force and the parameters used for their calculation are given in Table 1 and Table 2 respectively. Values for the surface potentials were interpolated from measurements in Figure 2.

For both cases, the van der Waals force is attractive and expressed as

$$\mathbf{F}_{vdW} = 2.03 \times 10^{-21} d_p^3 \frac{\alpha_{rtd}}{z^2(z+d_p)^2} \mathbf{n}_z \qquad (1,2)$$

The magnitude and sign of the three forces ($F_{el}$, $F_{vdW}$ and their sum $F_{DLVO}$) is shown in Table 1 for the two cases. For pH = 11.7, the repulsive electrostatic force is one order of magnitude higher than attractive van der Waals force while for pH = 1.1, the attractive van der Waals force is one order of magnitude larger than the repulsive electrostatic force. Thus, the force between the particle and the substrate is repulsive for pH = 11.7, and attractive for pH = 1.1. The final expression of the forces are given as follows for the two cases:

$$\mathbf{F}_{DLVO-} = \left[ -6.61 \times 10^{-11} \exp(-z/\kappa^{-1}) + 2.03 \times 10^{-21} d_p^3 \frac{\alpha_{rtd}}{z^2(z+d_p)^2} \right] \mathbf{n}_z, \text{ for pH} = 11.7 \qquad 13$$

$$\mathbf{F}_{DLVO+} = \left[ -8.24 \times 10^{-11} \exp(-z/\kappa^{-1}) + 2.03 \times 10^{-21} d_p^3 \frac{\alpha_{rtd}}{z^2(z+d_p)^2} \right] \mathbf{n}_z, \text{ for pH} = 1.1 \qquad 14$$

Figure 5a describes our numerical simulation of the evaporation of a 5 nL drop containing colloidal particles on a glass substrate at ambient temperature, with an initial pH of 11.7. The thermophysical properties and main parameters used in the simulation are given in Table 3 and Table 4 respectively. In Figure 5a, the evaporation time increases from top to bottom. Streamlines and velocity amplitude are shown left, and particle concentrations are shown right. Although the numerical code allows for receding, the wetting line is pinned during the entire drying process, because of the low receding angle value for water on glass ($\phi_{receding} \sim 1°$ in [66]), and because of the radial flow that bring particles to the wetting line [21]. This flow pattern arises because of the maximum evaporation flux at the wetting line. Because of the repulsive DLVO force between the particles and the substrate, the particles do not stick to the substrate but follow the main radial flow. The formation of a ring starts as early as $t = 1.8$ s when the particle concentration at the wetting line reaches 0.7 corresponding to maximum particle packing [67]. A view of the growing ring is shown in red from $t = 5.4$–$7.2$ s, with a skewed $z$- axis to better show the ring formation. Figure 6a compares the final deposit shapes obtained experimentally and numerically. The numerical profile corresponding to the lower value of the depinning angle ($\phi_{rec}$ = 25°[68], used throughout the simulation in Figure 5a) appears to match the experimental profile better than another simulation with larger depinning angle ($\phi_{rec}$ = 85°). In the latter case, depinning has occurred too abruptly, resulting in an inner slope of the ring that is steeper than in the measurement.

Comparison between the deposit profiles obtained by simulation and measurement for pH = 2.8 is shown in Figure 6b. The measured profile is more uniform than the deposit in Figure 6a and shows a peripheral ring along with a uniform deposit within the wetted area. The measured height of the ring is around twice of the height of the uniform deposit within the wetted area (Figure 6b). Due to the attractive DLVO force between the particles and the substrate at pH = 2.8 (see Table 1), a uniform layer of particles deposits at the bottom of the drop.

To increase the uniformity of the deposit profile over the entire wetted area, another experiment is performed in which a 4 nL water droplet at pH = 1.1 evaporates in a refrigerator with ambient temperature of 10.9°C, which corresponds to a final evaporation time of 16 s, almost twice the final evaporation time for the droplet at $T_{amb}$ =25.5°C. Figure 5b describes the corresponding numerical simulation, with parameters described in Table 4. Streamlines and velocity amplitudes are shown left, and particle concentrations are shown right. A similar radial flow pattern as in Figure 5a is observed, however with a lower intensity. In Figure 5b also, the DLVO force between the particles and the substrate is attractive (eq. 14, see Table 1). This causes a uniform layer of particles to be deposited at the bottom of the drop. Note that by lowering the ambient temperature the evaporation time is almost doubled. The corresponding reduction in evaporation flux inhibits the tendency of particles to flow to the wetting line, and allows more time for the particles to be attracted towards the substrate. This results in a more uniform deposit over the entire wetted area. Figure 6c compares the profile of the deposit measured by laser profilometer with the simulation results. The measurement shows a uniform deposit with average height of about 400 nm. The structure, shape and height of the deposit calculated numerically with $\phi_{rec} = 25°$ [68] reproduces this deposit structure qualitatively and quantitatively. The measured profile in Figure 6c shows a relatively uniform deposit, with some noise due to light scattering.

Simulations at intermediate pH value (pH = 5.8) could not reproduce the deposit profile given by the experiment. Experiment shows a deposit profile with flocculated particles (Figure 4c and Figure 3, measured deposit height ~ 0.5 μm) while the simulation predicts a ring pattern (height ~ 1.7 μm) with a very thin layer of deposit at the center (height ~ 0.05 micron). The reason for this discrepancy is proposed to be particle aggregation due to the dominance of attractive interparticle van der Waals forces (see Table 1), which is very reasonable given the low surface potential value. These aggregates will then tend to deposit over the entire area by sedimentation. As mentioned in section 3, this attractive force between the particles is not included in the numerical model.

### 4.3 Phase diagram

In this section, we propose a phase diagram for predicting the deposit shapes from drop containing colloidal particles observed in this paper and in our previous study [21]. The diagram in Figure 8 puts forth that the deposit shape results from the competition of three convective flow patterns, each inducing a specific deposit shape, as shown in Figure 7. The magnitude of each flow pattern can be estimated using analytical relations. The first flow pattern is the radial flow caused by the maximum evaporation rate at the pinned wetting line, and the corresponding deposit is a peripheral ring. An analytical expression for the radial velocity is provided by Hu and Larson [69], which scales as $V_{\text{rad}} \sim j/\rho$.

The second relevant flow pattern is the transport of particles normally towards the substrate, occurring in the case of an attractive DLVO force. The corresponding deposit pattern is a uniform layer with diameter equal to the initial wetted diameter. This velocity scales as in the equation 11 of the present paper: $V_{\text{DLVO+}} \sim 2F_{\text{DLVO}}/6\pi\mu d_{\text{p}}$, where $F_{\text{DLVO+}}$ is the magnitude of the attractive DLVO force calculated at the Debye length.

The third flow pattern is a Marangoni recirculation loop, and the corresponding deposit is a central bump, i.e. a hilly accumulation with a diameter much smaller than the initial wetted diameter of the drop. The typical loop velocity is given analytically by Hu and Larson [60] and scales as $V_{\text{Ma}} \sim (1/32)(\beta\phi_1^2 \Delta T/\mu)$. In this equation, $\phi$ is the wetting angle of the drop, $\mu$ is dynamic viscosity, $\beta$ is the gradient of surface tension with respect to the temperature, and $\Delta T$ is the temperature difference between the edge and the top of the droplet.

In Figure 8, we express the competition between these three convective flow patterns using a two-dimensional phase diagram. The horizontal axis expresses the ratio of the Marangoni recirculation over the radial flow, $V_{\text{Ma}}/V_{\text{rad}}$. The vertical axis expresses the ratio of the particle deposition driven by DLVO forces over the radial flow, $V_{\text{DLVO+}}/V_{\text{rad}}$. In Figure 8, five deposit patterns (A-E) obtained experimentally in this paper and in our previous study are shown for the three domains of the phase diagram. For a system where Marangoni effect and DLVO forces are negligible with respect to the radial flow caused by a maximum evaporation rate at the wetting line, the pattern is a ring (Figure 7(1)). This case corresponds to $V_{\text{Ma}}/V_{\text{rad}} = 0$ and $V_{\text{DLVO+}}/V_{\text{rad}} = 0$ and falls on the origin of the map shown in Figure 8 (case C). In a system where the attractive DLVO forces between particles and substrate dominate over radially outward flow and over Marangoni convection, the pattern is a uniform deposit over the entire wetted area (Figure 7(2)). The deposit is smooth if interparticle forces are negligible or is a uniform dispersion of flocculated particles if interparticle forces are important. Such system is shown as Case E in Figure 8 for the case $V_{\text{DLVO+}}/V_{\text{rad}} \sim 2.8 \times 10^5$. For a system where Marangoni convection dominates over radially outward flow and over attractive DLVO forces, the deposition pattern will be a central bump (Figure 7(3)). Such system is shown as Case D in Figure 8 for the case $V_{\text{Ma}}/V_{\text{rad}} \sim 34$. The phase diagram has therefore three domains, a bottom left corner where the deposit tends to be a ring, a top left zone where the deposit tends to be a uniform layer, and a bottom right zone where the deposit tends to be a central bump.

## 5 Conclusions

This paper describes the influence of the DLVO forces on the shape of deposits left by drying drop containing colloidal particles. Experiments are performed where the pH is used to control DLVO forces between the colloidal particles. An in-house numerical modeling able to simulate the evaporation of a drop containing colloidal particles is extended to consider the van der Waals and electrostatic forces between the particles and the deposit. Agreement between the simulations and the experiments is very good, with a discrepancy at intermediate pH that is attributed to the neglect of interparticle forces. A phase diagram is put forth to explain that the ratio of three characteristic velocities determines the deposit to be either a peripheral ring, a central bump and a uniform deposit.

## 6 Acknowledgements

The authors gratefully acknowledge financial support for this work from the Chemical Transport Systems Division of the US National Science Foundation through grant 0622849. We thank

# 7 Figures

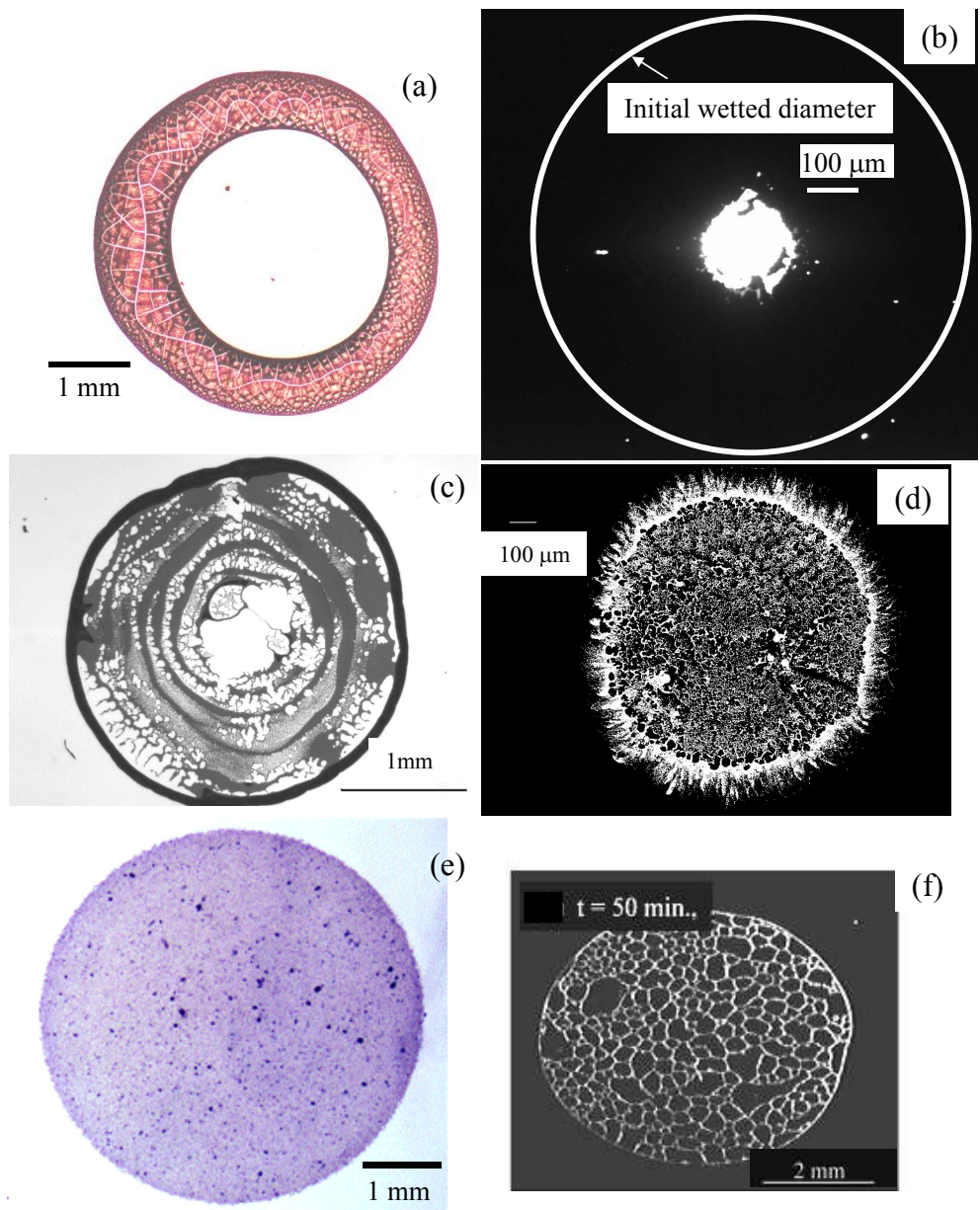

Figure 1: A multiplicity of deposits can be obtained after the drying of a drop containing colloidal particles: (a) ring-like pattern from an aqueous drop containing 60 nm polystyrene spheres on titanium substrate [30] (with permission from ACS); (b) central bump obtained after the drying of a 38 nL isopropanol drop on PDMS at ambient temperature [21]; (c) multiple rings from a µL water drop containing 1 µm polystyrene microspheres on glass (our work); (d) fingering at wetting line obtained from a µL isopropanol drop with 1µm polystyrene microspheres on glass (our work); (e) uniform deposition pattern of 60 nm hydroxyapatite particles from aqueous drop on titanium disk [33] (with permission from ACS); (f) hexagonal cells from surfactant-laden aqueous drop containing polystyrene microspheres on hydrophobic OctadecylTricholoroSilane (OTS) substrate [24] (with permission from ACS).

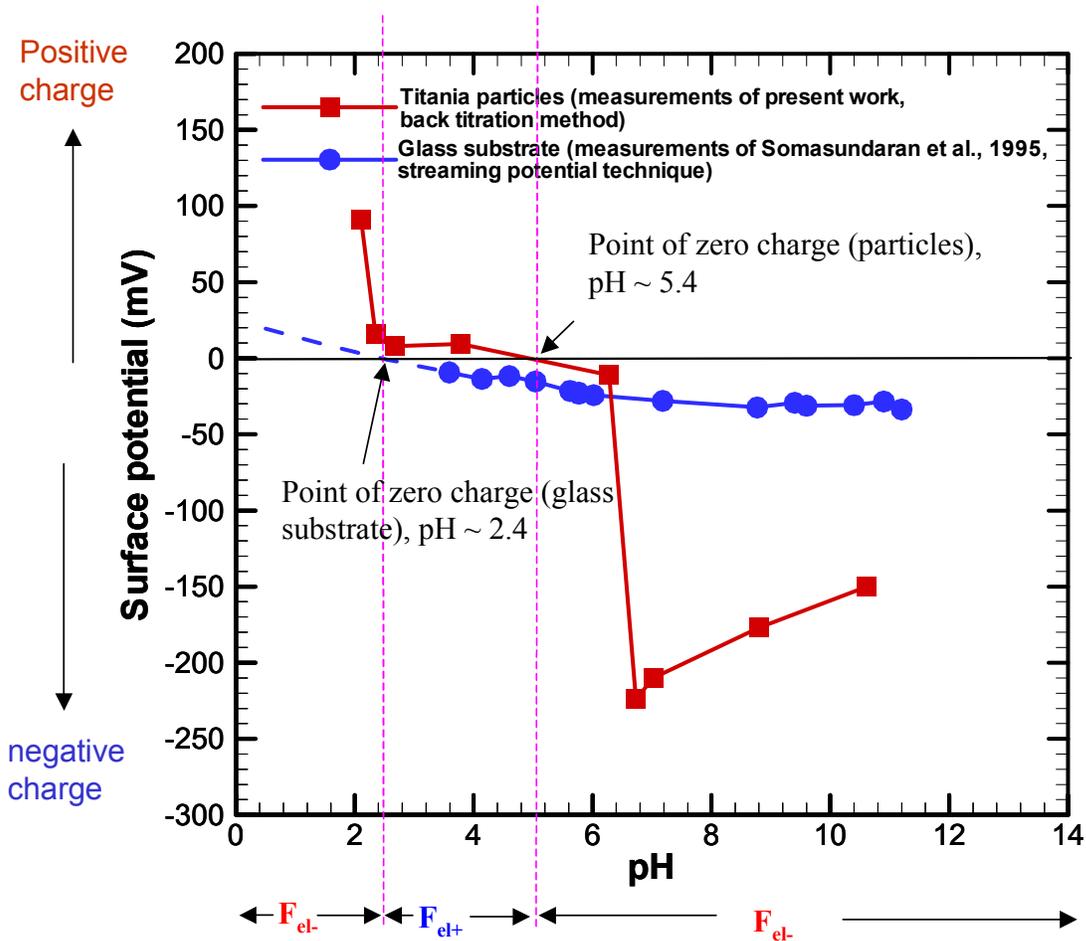

Figure 2: The measured surface potential as a function of pH for 25 nm Titania particles in water, and for a glass substrate. The surface potential values for the substrate are taken from the work of Somasundaran et al. [49].

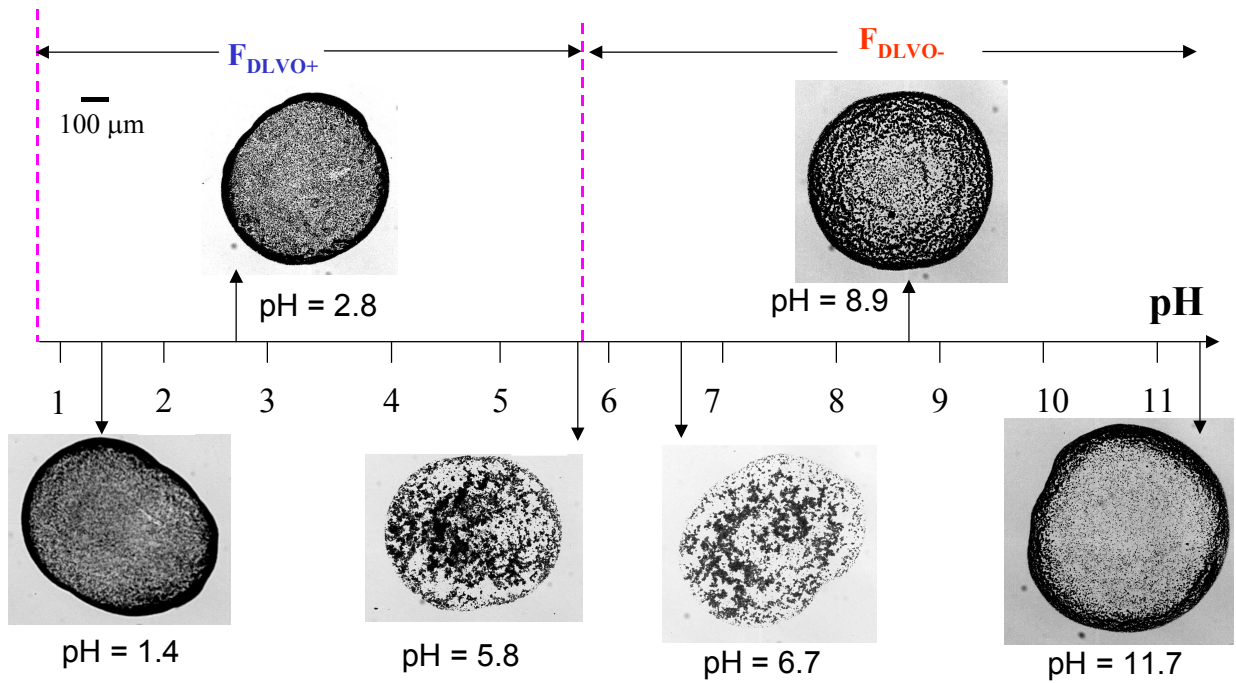

Figure 3: Deposit patterns obtained for different values of pH for a water-glass-Titania system. The diameter and volume concentration of the particles are respectively 25 nm and 2%. See associated movies at pH = 11.7 [44], pH = 5.8 [65] and pH = 2.8 [64].

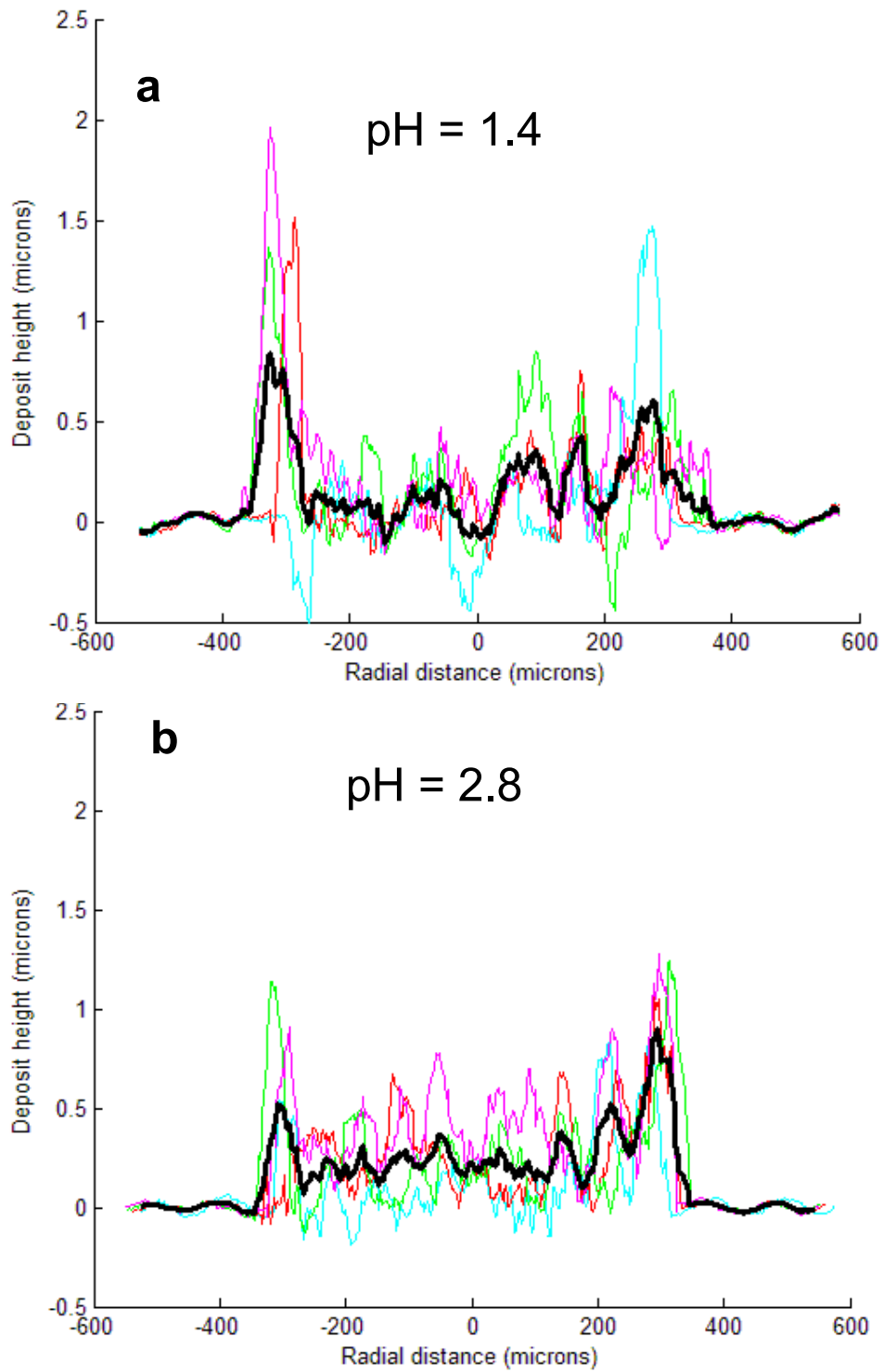

Figure 4: Deposit profiles obtained using a laser profilometer for different pH values. Colored profiles are measurements along four azimuthal angles, and the bold black line is their average.

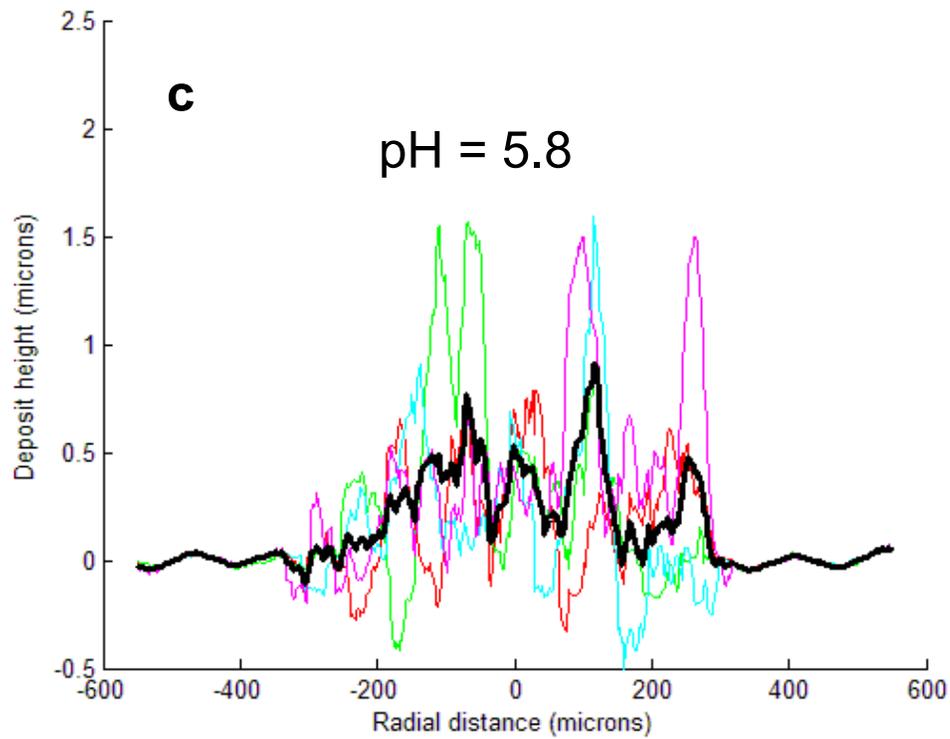

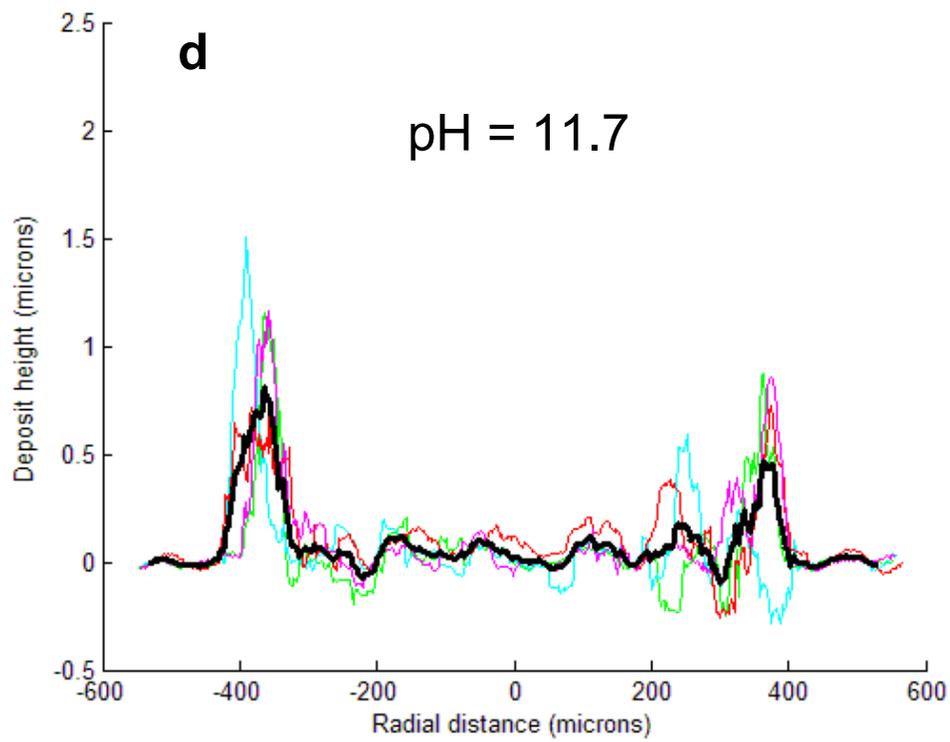

Figure 4: continued

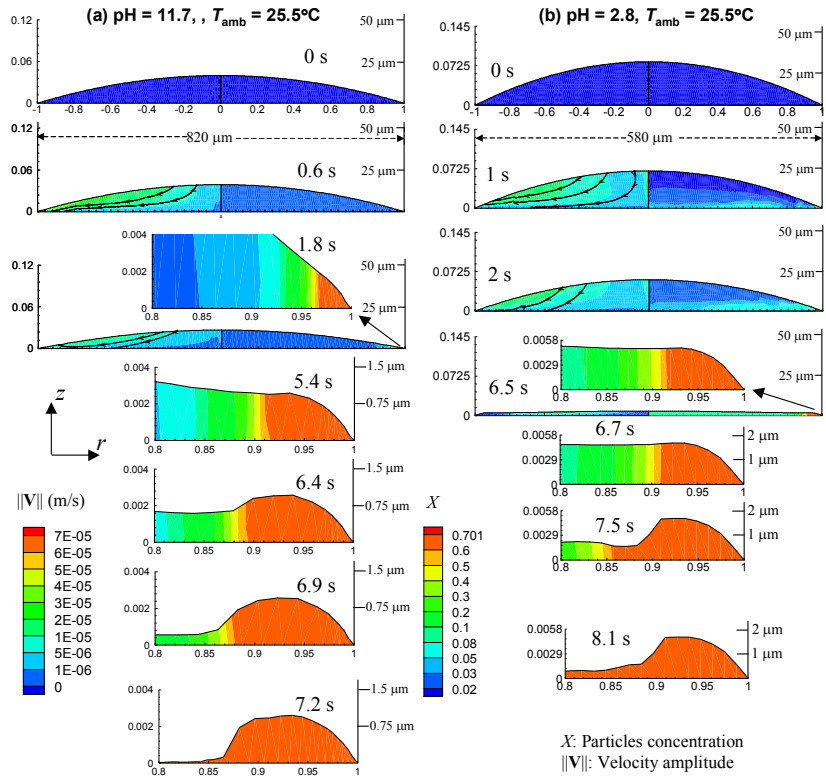
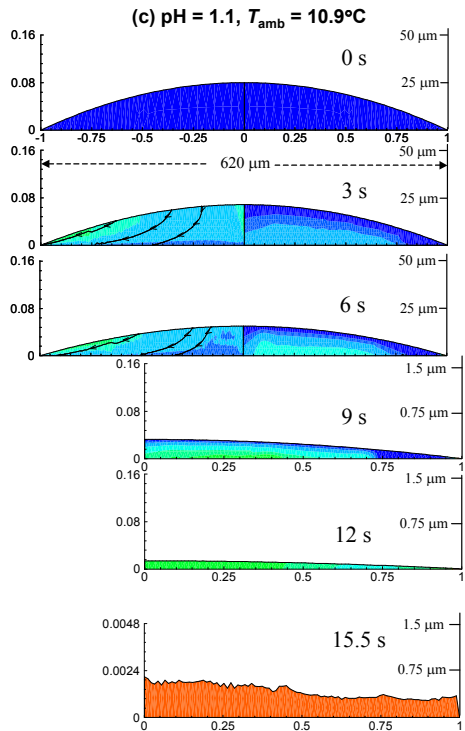

Figure 5: Simulations of the evaporation of nanoliter colloidal water drops on glass: (a) ring formation for pH = 11.7, $T_{amb}$=25.5°C; (b) deposit with ring formation for pH = 2.8, $T_{amb}$=25.5°C; (c) uniform deposit formation for pH = 1.1, $T_{amb}$=10.9°C.. Particle concentration contours (left) and streamlines superposed to velocity amplitude (right) are shown. A deposit starts forming when the particle concentration reaches 0.7 (in red). Note that the aspect ratio is increased towards the vertical axis in the last stages of the evaporation.

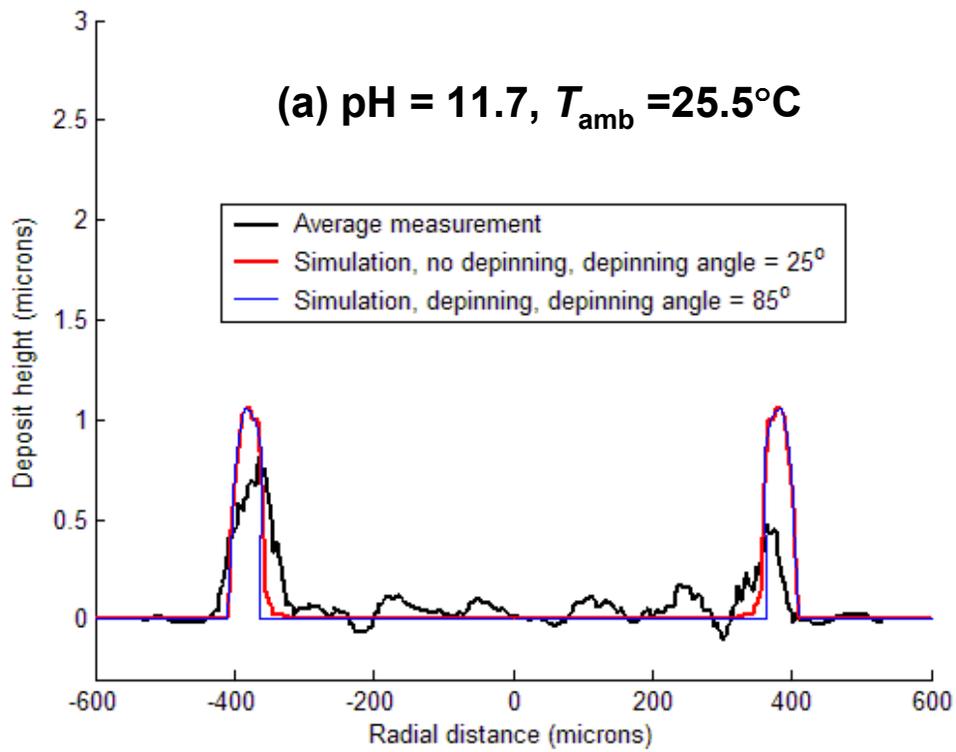
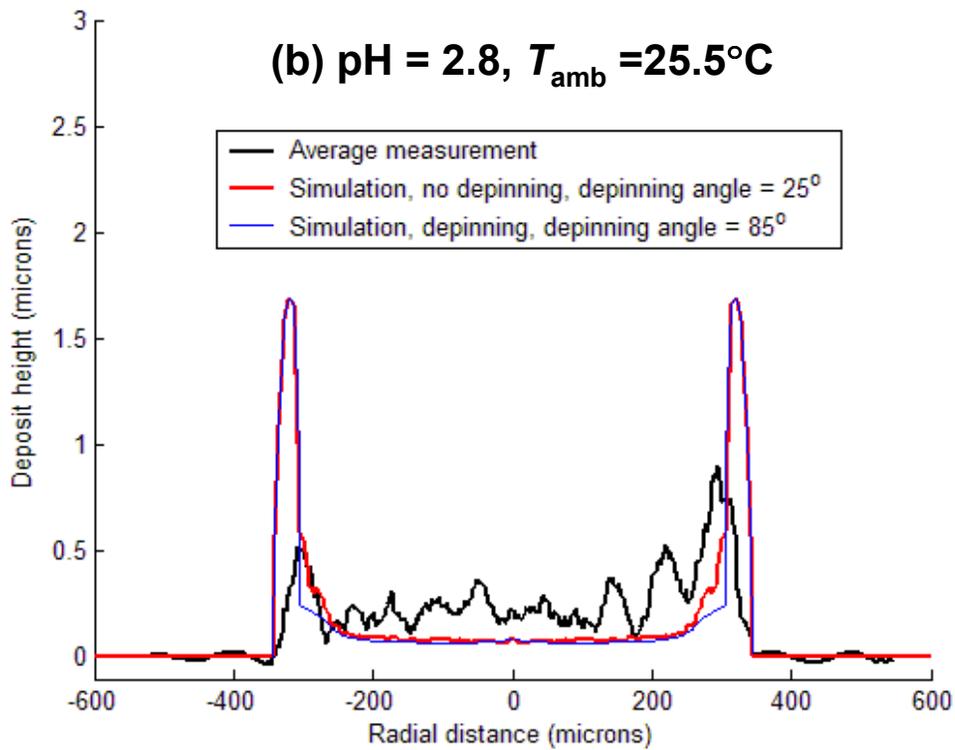

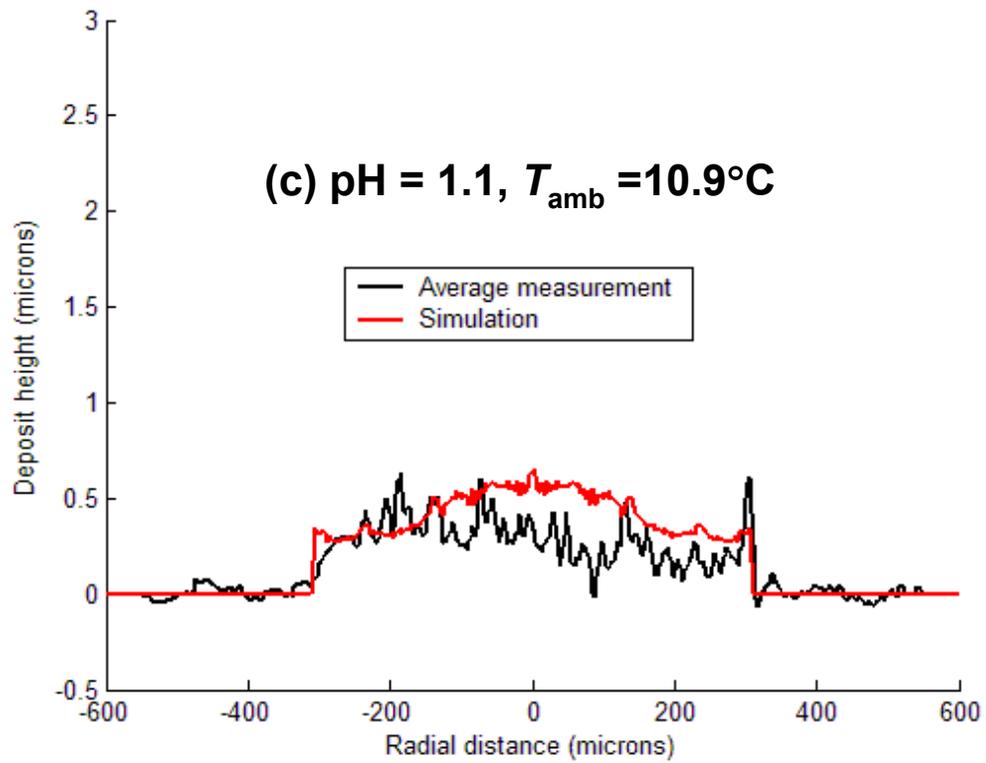

Figure 6: Comparisons between measured and simulated deposit profiles for (a) pH = 11.7, $T_{amb}$=25.5 °C; (b) pH = 2.8, $T_{amb}$=25.5°C and (c) pH = 1.1, $T_{amb}$=10.9°C.

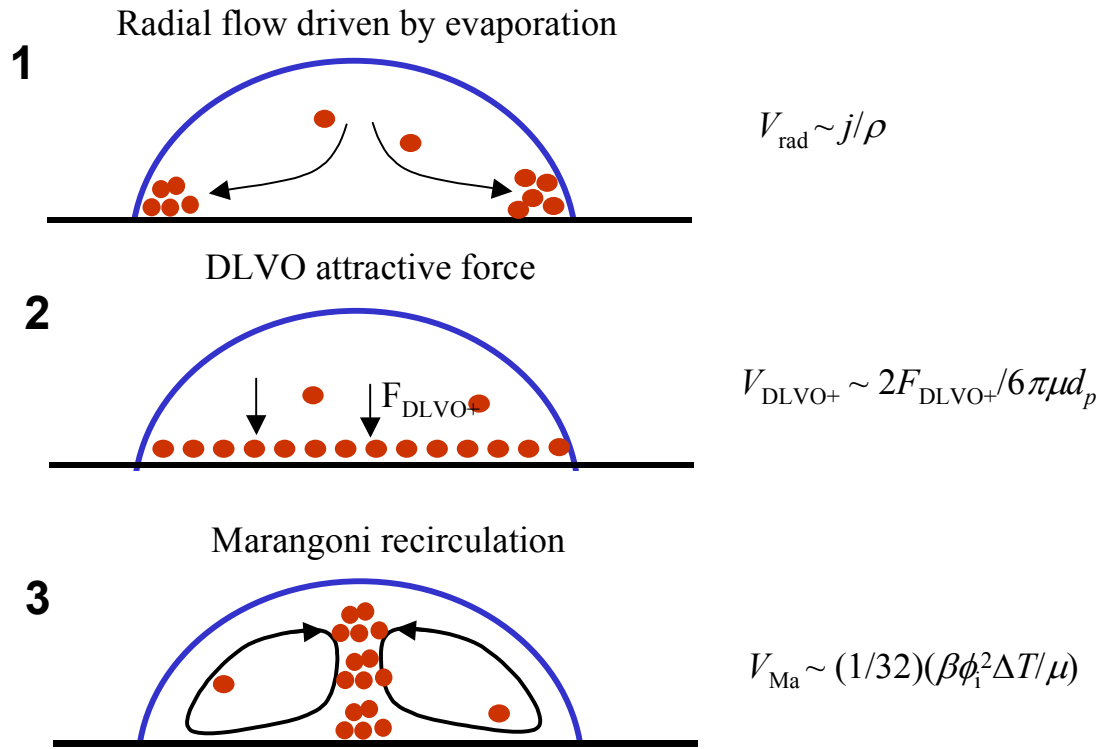

Figure 7: Three convective mechanisms compete to form the deposit. In (1) a ring forms due to radial flow caused by a maximum evaporation rate at the pinned wetting line; in (2) a uniform deposit forms due to an attractive DLVO force between the particles and the substrate; in (3) a central bump forms due to a Marangoni recirculation loop.

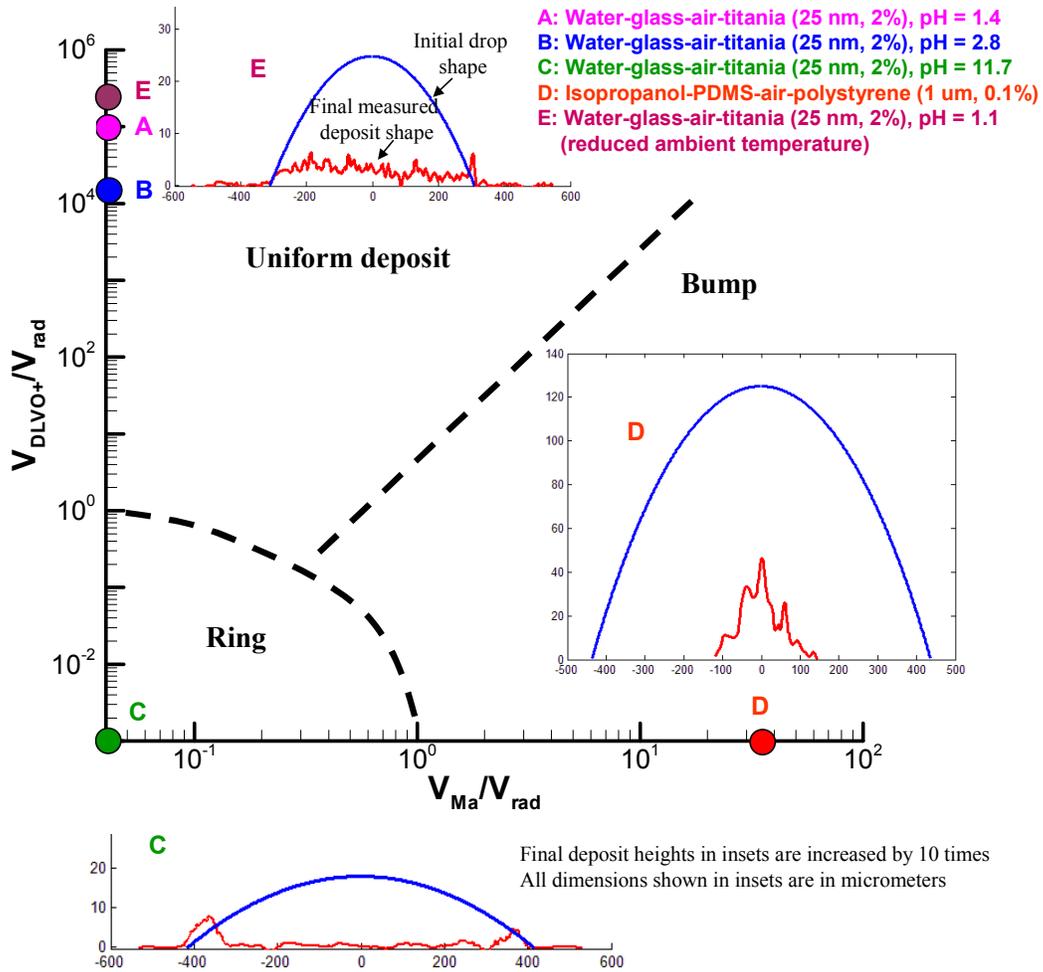

Figure 8: Phase diagram for self-assembly of nanoparticles during drop drying on a solid surface. The ratio of three characteristic velocities ($V_{rad}$, $V_{DLVO+}$, and $V_{Ma}$) determines the final pattern shape. $V_{rad}$ is the radial flow velocity scale caused by the maximum evaporation rate at the pinned wetting line, $V_{DLVO+}$ is the velocity scale caused by an attractive DLVO force and $V_{Ma}$ is the Marangoni velocity scale. Cases A, B, C and E are experiments performed in this work, while case D is an experiment performed in [21].

# 8 Tables

Table 1: Value of the forces for different pH cases. Forces are estimated at the Debye length, except for the interparticle force, which is estimated at twice that length.

| pH | $\kappa^{-1}$ (nm) | $a$ (in eq. 4) | $F_{el}$ (N) | $F_{vdW}$ (N) | $F_{DLVO}$ $(=F_{el}+F_{vdW})$ (N) | Nature of $F_{DLVO}$ | $F_{vdWp}$ (N) |
|---|---|---|---|---|---|---|---|
| 1.1 | 1.01 | $-8.24 \times 10^{-11}$ | $-3.04 \times 10^{-11}$ | $1.59 \times 10^{-10}$ | $1.29 \times 10^{-10}$ | Attractive | $6.20 \times 10^{-12}$ |
| 1.4 | 1.16 | $-2.18 \times 10^{-11}$ | $-8.02 \times 10^{-12}$ | $1.10 \times 10^{-10}$ | $1.02 \times 10^{-10}$ | Attractive | $4.7 \times 10^{-12}$ |
| 2.8 | 2.42 | $1.83 \times 10^{-13}$ | $6.73 \times 10^{-14}$ | $1.43 \times 10^{-11}$ | $1.43 \times 10^{-11}$ | Attractive | $1.08 \times 10^{-12}$ |
| 5.8 | 18.34 | $-5.69 \times 10^{-15}$ | $-2.09 \times 10^{-15}$ | $6.97 \times 10^{-15}$ | $4.88 \times 10^{-15}$ | Attractive | $8.58 \times 10^{-15}$ |
| 6.7 | 931.7 | $-6.76 \times 10^{-13}$ | $-2.48 \times 10^{-13}$ | $2.14 \times 10^{-21}$ | $-2.48 \times 10^{-13}$ | Repulsive | $7.28 \times 10^{-18}$ |
| 8.9 | 22.12 | $-4.73 \times 10^{-13}$ | $-1.74 \times 10^{-13}$ | $1.43 \times 10^{-14}$ | $-1.60 \times 10^{-13}$ | Repulsive | $1.29 \times 10^{-14}$ |
| 11.7 | 3.46 | $-6.61 \times 10^{-11}$ | $-2.43 \times 10^{-11}$ | $5.21 \times 10^{-12}$ | $-1.90 \times 10^{-11}$ | Repulsive | $5.28 \times 10^{-13}$ |

Table 2: Experimental quantities for two pH cases

| Solution | pH | $\kappa^{-1}$ (nm) | [H+] (mol/L) | [OH-] (mol/L) | [Na+] (mol/L) | [Cl-] (mol/L) | $n$ | $\sigma_p$ [Cm$^{-2}$] | $\psi_p$ [mV] | $\zeta_s$ [mV] |
|---|---|---|---|---|---|---|---|---|---|---|
| 1 | 11.7 | 3.46 | $1.82 \times 10^{-12}$ | $5.5 \times 10^{-3}$ | $5.5 \times 10^{-3}$ | 0 | $3.3 \times 10^{24}$ | -0.08 | -141 | -34 |
| 2 | 1.1 | 1.01 | $7.94 \times 10^{-2}$ | $1.26 \times 10^{-13}$ | 0 | $7.94 \times 10^{-2}$ | $4.78 \times 10^{25}$ | 0.078 | 77 | 13 |

Table 3: Thermophysical properties used in the simulations at 25°C [70]

| Substance | Density [kgm$^{-3}$] | Thermal conductivity [Wm$^{-1}$K$^{-1}$] | Specific heat [Jkg$^{-1}$K$^{-1}$] | Viscosity [Pa-s] | Surface energy [Jm$^{-2}$] | Latent heat [Jkg$^{-1}$] |
|---|---|---|---|---|---|---|
| Water | 997 | 0.607 | 4180 | 9.0e-4 | 7.2e-2 | 2445e3 |
| Glass | 2200 | 1.38 | 740 | - | - | - |

Table 4: Parameters used in the simulations for reproducing experiments

| Case | Particles Concentration, $X$ (v/v) | $\kappa^{-1}$ (nm) | $r_{max}$ μm | $V_i$ nL | $\phi_i$ | $T_{amb}$ | $H$ | $\gamma = f(T)$ | $\mu = f(T)$ |
|---|---|---|---|---|---|---|---|---|---|
| pH = 11.7 | 2% | 3.46 | 410 | 5 | 5° | 25.5°C | 45.2% | No | Yes |
| pH = 1.1 | 2% | 1.01 | 310 | 4 | 9° | 10.9°C | 44.0% | No | Yes |

# 9  References

[1] Hiemenz, P. C. *Principles of Colloid and Surface Chemistry*. Marcel Dekker, New York, 1986.

[2] Heim, T., Preuss, S., Gerstmayer, B., Bosio, A. and Blossey, R. Deposition from a drop: morphologies of unspecifically bound DNA. *Journal of Physics: Condensed Matter*, 17, (2005), S703-S716.

[3] Wang, D., Liu, S., Trummer, B. J., Deng, C. and Wang, A. Carbohydrate microarrays for the recognition of cross reactive molecular markers of microbes and host cells. *Nature biotechnology*, 20, (2002), 275-281.

[4] Dugas, V., Broutin, J. and Souteyrand, E. Droplet evaporation study applied to DNA chip manufacturing. *Langmuir*, 21, (2005), 9130-9136.

[5] Carroll, G. T., Wang, D., Turro, N. J. and Koberstein, J. T. Photochemical micropatterning of carbohydrates on a surface. *Langmuir*, 22, (2006), 2899-2905.

[6] Smalyukh, I. I., Zribi, O. V., Butler, J. C., Lavrentovich, O. D. and Wong, G. C. L. Structure and dynamics of liquid crystalline pattern formation in drying droplets of DNA. *Physical Review Letters*, 96, (2006), 177801.

[7] Blossey, R. and Bosio, A. Contact line deposits on cDNA microarrays: A 'Twin-spot effect'. *Langmuir*, 18, (2002), 2952-2954.

[8] Larson, R. G., Perkins, T. T., Smith, D. E. and Chu, S. Hydrodynamics of a DNA molecule in a flow field. *Physical Review E*, 55, (1997), 1794.

[9] Kralchevsky, P. A. and Denkov, N. D. Capillary forces and structuring in layers of colloidal particles. *Current opinion in Collidal and interface science*, 6, (2001), 383-201.

[10] Dushkin, C. D., Lazarov, G. S., Kotsev, S. N., Yoshimura, H. and Nagayama, K. Effect of growth conditions on the structure of two-dimensional latex crystals: Experiment. *Colloid Polymer Science*, 277, (1999), 914-930.

[11] Dushkin, C. D., Yoshimura, H. and Nagayama, K. Nucleation and growth of two-dimensional colloidal crystals. *Chemical Physics Letters*, 204, (1993), 455-460.

[12] Denkov, N. D., Velvev, O. D., Kralchevsky, P. A., Ivanov, I. B., Yoshimura, H. and Nagayama, K. Mechanism of Formation of Two-Dimensional Crystals from Latex Particles on the substrates. *Langmuir*, 8, (1992), 3183-3190.

[13] Denkov, N. D., Velvev, O. D., Kralchevsky, P. A. and Ivanov, I. B. Two-dimensional crystallization. *Nature*, 361, (1993), 26.

[14] Abkarian, M., Nunes, J. and Stone, H. A. Colloidal crystallization and Banding in a Cylindrical Geometry. *Journal of American Chemical Society*, 126, (2004), 5978-5979.

[15] Subramaniam, A. B., Abkarian, M. and Stone, H. A. Controlled assembly of jammed colloidal shells on fluid droplets. *Nature materials*, 4, (2005), 553-556.

[16] Cuk, T., Troian, S. M., Hong, C. M. and Wagner, S. Using convective flow splitting for the direct printing of fine copper lines. *Applied Physics Letters*, 77, (2000), 2063.

[17] Ondarcuhu, T. and Joachim, C. Drawing a single nanofibre over hundreds of microns. *Europhysics Letters*, 42, (1998), 215-220.

[18] Maillard, M., Motte, L. and Pileni, M.-P. Rings and Hexagons made of nanocrystals. *Advanced Materials*, 13, (2001), 200-204.

[19] Personal communication with Greg Gillen at National Institute of Standards and Technology (NIST), (2005).

[20] Deegan, R. D., Bakajin, O., Dupont, T. F., Huber, G., Nagel, S. R. and Witten, T. A. Capillary flow as the cause of ring stains from dried liquid drops. *Nature*, 389, (1997), 827-829.

[21] Bhardwaj, R., Fang, X. and Attinger, D. Pattern formation during the evaporation of a colloidal nanoliter drop: a numerical and experimental study. *New Journal of Physics*, 11, (2009), 075020.

[22] Sommer, A. P. and Franke, R. Biomimicry patterns with Nanosphere Suspensions. *Nano Letters*, 3, (2003), 573.

[23] Hu, H. and Larson, R. G. Marangoni Effect Reversed Coffee-Ring Depositions. *Journal of Physical Chemistry B*, 110, (2006), 7090-7094.

[24] Truskett, V. N. and Stebe, K. J. Influence of Surfactants on an Evaporating Drop: Fluorescence Images and Particle Deposition Patterns. *Langmuir*, 19, (2003), 8271-8279.

[25] Deegan, R. D. Pattern Formation in Drying Drops. *Physical Review E*, 61, (2000), 475-485.

[26] Deegan, R. D., Bakajin, O., Dupont, T. F., Huber, G., Nagel, S. R. and Witten, T. A. Contact line deposits in an evaporating drop. *Physical Review E*, 62, (2000), 756.

[27] Ristenpart, W. D., Kim, P. G., Domingues, C., Wan, J. and Stone, H. A. Influence of Substrate Conductivity on Circulation Reversal in Evaporating Drops. *Physical Review Letters*, 99, (2007), 234502.

[28] Lim, J. A., Lee, W. H., Lee, H. S., Lee, J. H., Park, Y. D. and Cho, K. Self-organization of ink-jet-printed triisopropylsilylethynyl pentacene via Evaporation-induced flows in a drying droplet. *Advanced functional materials*, 18, (2008), 229-234.
[29] Private communication with Greg Gillen at National Institute of Standards and Technology (NIST), (2005).
[30] Sommer, A. P., Ben-Moshe, M. and Magdassi, S. Self-Discriminative Self-Assembly of Nanospheres in Evaporating Drops. *Journal of Physical Chemistry B*, 108, (2004), 8-10.
[31] Park, J. and Moon, J. Control of Colloidal Particle Deposit Patterns within Pioliter Droplets Ejected by Inkjet printing. *Langmuir*, 22, (2006), 3506-3513.
[32] Sommer, A. P., Cehreli, M., Akca, K., Sirin, T. and Piskin, E. Superadhesion: Attachment of Nanobacteria to Tissues - Model Simulation. *Crystal Growth and Design*, 5, (2005), 21-23.
[33] Sommer, A. P. Suffocation of Nerve Fibers by Living Nanovesicles: A Model Simulation - Part II. *Journal of Proteome Research*, 3, (2004), 1086.
[34] Yan, Q., Gao, L., Sharma, V., Chiang, Y.-M. and Wong, C. C. Particle and substrate charge effects on colloidal self-assembly in a sessile drop. *Langmuir*, 24, (2008), 11518-11512.
[35] Andreeva, L. V., Koshkin, A. V., Lebedev-Stepanov, P. V., A N Petrov and Alfimov, M. V. Driving forces of the solute self-organization in an evaporating liquid droplet. *Colloids and Surfaces A: Physicochem. Eng. Aspects*, 300, (2007), 300-306.
[36] Onoda, G. and Somasundaran, P. Two- and One-Dimensional Flocculation of Silica Spheres on Substrates. *Journal of Colloid and Interface Science*, 118, (1987), 169-175.
[37] Kralchevsky, P. A., Denkov, N. D., Paunov, V. N., Velvev, O. D., Ivanov, I. B., Yoshimura, H. and Nagayama, K. Formation of two-dimensional colloid crystals in liquid films under the action of capillary forces. *Journal of Physics: Condensed Matter*, 6, (1994), A395-A402.
[38] Widjaja, E. and Harris, M. T. Particle deposition study during sessile drop evaporation. *AIChe journal*, 54, (2008), 2250-2260.
[39] Petsi, A. J. and Burganos, V. N. Evaporation-induced flow in an inviscid liquid line at any contact angle. *Physical Review E*, 73, (2006), 041201.
[40] Uno, K., Hayashi, K., Hayashi, T., Ito, K. and Kitano, H. Partticle adsorption in evaporating droplets of polymer latex dispersions on hydrophobic and hydrophillic surfaces. *Colloid and Polymer Science*, 276, (1998), 810.
[41] Kosmulski, M. *Chemical properties of Materials Surfaces*, New York, 2001.
[42] Morrison, I. D. and Ross, S. *Colloidal Dispersions, Suspensions, Emulsions, and Foams*. John Wiley and Sons, Inc, New York, 2002.
[43] Barbulovic-Nad, I., Lucente, M., Sun, Y., Zhang, M., Wheeler, A. R. and Bussmann, M. Bio-Microarray Fabrication Techniques- A Review. *Critical Reviews in Biotechnology*, 26, (2006), 237-259.
[44] Multimedia provided with manuscript (filename = ring.avi), Available at www.me.columbia.edu/lmtp/Langmuir_movies/ring.avi
[45] Le, F.-I., Leo, P. H. and Barnard, J. A. Dendrimer pattern formation in evaporationg drops: Solvent, size and concentration effects. *Journal of Physical Chemistry C*, 112, (2008), 14266-14273.
[46] Le, F.-I., Leo, P. H. and Barnard, J. A. Temperature-dependent formation of dendrimer islands from ring structures. *Journal of Physical Chemistry B*, 112, (2008), 16497–16504.
[47] Somasundaran, P. *Introduction to surface and colloid chemistry*. Solloid publishers, New York, 2006.
[48] Kobayashi, M., nanaumi, H. and Muto, Y. Initial deposition rate of latex particles in the packed bed of zirconia beads. *Colloids and Surfaces A: Physicochem. Eng. Aspects*, 347, (2009), 2-7.
[49] Somasundaran, P., Shrotri, S. and Ananthapadmanabhan, K. P. *Deposition of latex particles: Theoretical and Experimental aspects*. Allied Publishers, City, 1995.
[50] Lameiras, F. S., Souza, A. L. D., Melo, V. A. R. D., Nunes, E. H. M. and Braga, I. D. Measurement of zeta potential of planar surfaces with a rotating disk. *Materials Research*, 11, (2008), 217-219.
[51] Tien, C. *Granular Filtration of Aerosols and Hydrosols*. Butterworths, Boston, 1989.
[52] Waldvogel, J. M. and Poulikakos, D. Solidification Phenomena in Picoliter Size Solder Droplet Deposition on a Composite Substrate. *International Journal of Heat and Mass transfer*, 40, (1997), 295-309.
[53] Haferl, S., Zhao, Z., Giannakouros, J., Attinger, D. and Poulikakos, D. *Transport Phenomena in the Impact of a Molten Droplet on a Surface: Macroscopic Phenomenology and Microscopic Considerations. Part I: Fluid Dynamics*. Begell House, NY, City, 2000.
[54] Attinger, D., Haferl, S., Zhao, Z. and Poulikakos, D. Transport Phenomena in the Impact of a Molten Droplet on a Surface: Macroscopic Phenomenology and Microscopic Considerations. Part II: Heat Transfer and Solidification. *Annual Review of Heat Transfer*, XI, (2000), 65-143.

[55] Bhardwaj, R., Longtin, J. P. and Attinger, D. A numerical investigation on the influence of liquid properties and interfacial heat transfer during microdroplet deposition onto a glass substrate. *International Journal of Heat and Mass Transfer*, 50, (2007), 2912-2923.

[56] Attinger, D. and Poulikakos, D. On Quantifying Interfacial Thermal and Surface Energy during Molten Microdroplet Surface Deposition. *Journal of Atomization and Spray*, 13, (2003), 309-319.

[57] Attinger, D. and Poulikakos, D. Melting and Resolidification of a Substrate caused by Molten Microdroplet Impact. *Journal of Heat Transfer*, 123, (2001), 1110-1122.

[58] Bhardwaj, R. and Attinger, D. Non-isothermal wetting during impact of millimeter size water drop on a flat substrate: numerical investigation and comparison with high speed visualization experiments. *International Journal of Heat and Fluid Flow*, 29, (2008), 1422-1435.

[59] Fukai, J., Zhao, Z., Poulikakos, D., Megaridis, C. M. and Miyatake, O. Modeling of the Deformation of a Liquid Droplet Impinging upon a Flat Surface. *Physics of Fluids A*, 5, (1993), 2588-2599.

[60] Hu, H. and Larson, R. G. Analysis of the Effects of Marangoni stresses on the Microflow in an Evaporating Sessile Droplet. *Langmuir*, 21, (2005), 3972-3980.

[61] Ward, C. A. and Stanga, D. Interfacial conditions during evaporation or condensation of water. *Phtsical Review E*, 64, (2001), 51509.

[62] Private communication with Prof Howard Stone at Princeton university, (2009).

[63] Bird, R. B., Stewart, W. E. and Lightfoot, E. N. *Transport Phenomena*. John Wiley & Sons, New York, 2007.

[64] Multimedia provided with manuscript (filename = ring_deposit.avi), Available at www.me.columbia.edu/lmtp/Langmuir_movies/ring_deposit.avi

[65] Multimedia provided with manuscript (filename = aggregates.avi), Available at www.me.columbia.edu/lmtp/Langmuir_movies/aggregates.avi

[66] Hu, H. and Larson, R. G. Evaporation of a sessile droplet on a substrate. *Journal of Physical Chemistry B*, 106, (2002), 1334-1344.

[67] Krishna, P. and Pandey, D. *Close-Packed Structures*. International Union of Crystallography University College Cardiff Press, Wales, 2001.

[68] Grigoriev, S. A., Millet, P. A., volobuev, S. A. and Fateev, V. N. Optimization of porous current collectors for PEM water electrolysers. *International Journal of Hydrogen Energy*, in press, (2009).

[69] Hu, H. and Larson, R. G. Analysis of Microfluid flow in an Evaporating sessile droplet. *Langmuir*, 21, (2005), 3963-3971.

[70] Lide, D. R. *CRC Handbook of Chemistry and Physics on CD-ROM*. Chapman and Hall/CRC, 2001.